\documentclass[]{spie}  
\usepackage[]{graphicx}
\usepackage{amssymb, amsmath}

\newcommand{\lya}{Ly$\alpha$}
\newcommand{\arcmin}{$^{\prime}$}
\newcommand{\arcsec}{$^{\prime\prime}$}
\newcommand{\degree}{$^{\circ}$}

\title{Methods for evaluating the performance of volume phase holographic gratings for the VIRUS spectrograph array} 

\author{Taylor S. Chonis\supit{a}, Gary J. Hill\supit{b},  J. Christopher Clemens\supit{c}, Bart Dunlap\supit{c}, Hanshin Lee\supit{b}
\skiplinehalf
\supit{a}The University of Texas at Austin, Department of Astronomy, 2515 Speedway, Stop C1400, Austin, TX, USA 78712; \\
\supit{b}The University of Texas at Austin, McDonald Observatory, 2515 Speedway, Stop C1402, Austin, TX, USA 78712; \\
\supit{c}University of North Carolina, Department of Physics \& Astronomy, CB3255, Chapel Hill, NC, USA 27599; \\
}

\authorinfo{Further author information: (Send correspondence to T.S.C.)\\T.S.C.: E-mail: tschonis@astro.as.utexas.edu, Telephone: 1-512-471-1495}

\begin{document} 
  \maketitle 

\begin{abstract}
The Visible Integral field Replicable Unit Spectrograph (VIRUS) is an array of at least 150 copies of a simple, fiber-fed integral field spectrograph that will be deployed on the Hobby-Eberly Telescope (HET) to carry out the HET Dark Energy Experiment (HETDEX). Each spectrograph contains a volume phase holographic grating as its dispersing element that is used in first order for $350 < \lambda \mathrm{(nm)} < 550$. We discuss the test methods used to evaluate the performance of the prototype gratings, which have aided in modifying the fabrication prescription for achieving the specified batch diffraction efficiency required for HETDEX. In particular, we discuss tests in which we measure the diffraction efficiency at the nominal grating angle of incidence in VIRUS for all orders accessible to our test bench that are allowed by the grating equation. For select gratings, these tests have allowed us to account for $>90$\% of the incident light for wavelengths within the spectral coverage of VIRUS. The remaining light that is unaccounted for is likely being diffracted into reflective orders or being absorbed or scattered within the grating layer (for bluer wavelengths especially, the latter term may dominate the others). Finally, we discuss an apparatus that will be used to quickly verify the first order diffraction efficiency specification for the batch of at least 150 VIRUS production gratings.  
\end{abstract}


\section{INTRODUCTION} \label{sec:intro}  
\begin{figure}[t]
   \begin{center}
   \begin{tabular}{c}
   \includegraphics[width=0.95\textwidth]{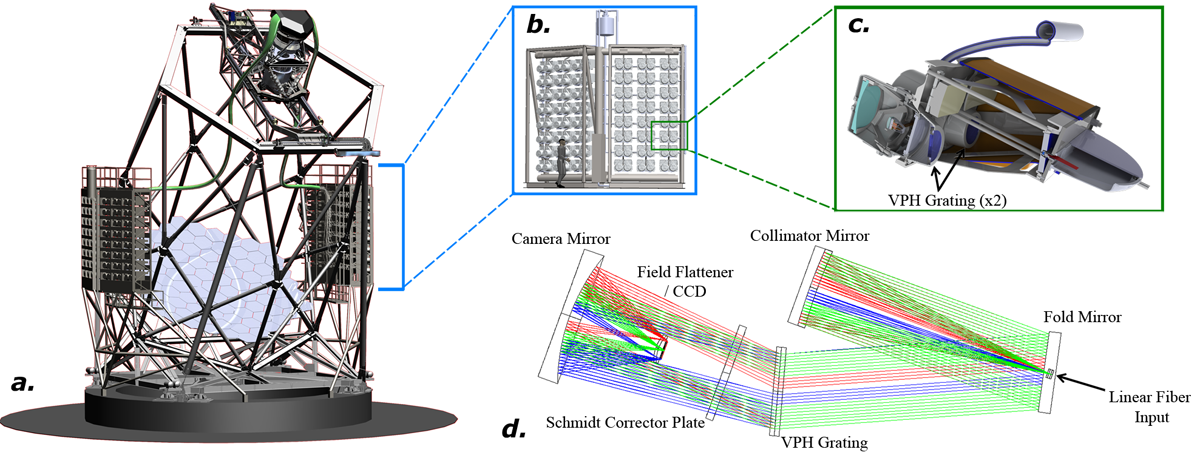}
   \end{tabular}
   \end{center}
   \caption[example] 
   { \label{fig:VIRUS} 
\textit{a}) A rendering of the upgraded HET showing the large enclosures mounted on either side of the telescope structure that contain VIRUS spectrographs. The green cables extending from the prime-focus instrument package to the enclosures are large bundles of fiber optics. \textit{b}) Close view of two enclosures, each containing an 8$\times$3 array of VIRUS units (48 total spectrographs). \textit{c}) Section view of a single VIRUS pair. \textit{d}) A ray trace of the VIRUS optical system.}
   \end{figure}

The upcoming Hobby-Eberly Telescope Dark Energy eXperiment (HETDEX; Ref. \citenum{Hill08a}) will amass a sample of $\sim0.8$ million \lya\ emitting galaxies (LAE) to be used as tracers of large-scale structure for constraining dark energy and measuring its possible evolution from $1.9 < z < 3.5$. To carry out the 120 night blind spectroscopic survey covering a 420 square degree field (9 Gpc$^{3}$), a revolutionary new multiplexed instrument called the Visible Integral field Replicable Unit Spectrograph (VIRUS; Ref. \citenum{Hill12a}) is being constructed\cite{Tuttle12} for the upgraded 9.2 m Hobby-Eberly Telescope (HET\footnote{The Hobby-Eberly Telescope is operated by McDonald Observatory on behalf of the University of Texas at Austin, the Pennsylvania State University, Ludwig-Maximillians-Universit\"{a}t M\"{u}nchen, and Georg-August-Universit\"{a}t Goettingen.}; Ref. \citenum{Hill12b}). The VIRUS spectrograph array consists of at least 150 copies (with a goal of 192) of a simple fiber-fed integral field spectrograph and for the first time has introduced industrial-scale replication to optical astronomical instrumentation. The spectrographs are mechanically built into unit pairs and are fed by dense-pack fiber bundle integral field units (IFU) with 1/3 fill factor, each consisting of 448 fiber optic elements with a core diameter of 266 $\mu$m (1.5\arcsec\ on the sky). Thus, each spectrograph images 224 fibers. At least 75 IFUs will be arrayed on the 22\arcmin\ diameter focal plane of the upgraded HET, yielding $\sim33000$ individual spectra per exposure. Each spectrograph consists of a double-Schmidt optical design with a volume phase holographic (VPH) diffraction grating at the pupil between a $f$/3.33 folded collimator and a $f$/1.25 cryogenic camera. The spectral coverage of VIRUS is $350 < \lambda \mathrm{(nm)} < 550$ at $R = \lambda / \Delta\lambda \approx 700$ for measuring the baryonic acoustic oscillation via the \lya\ emission of star-forming galaxies from $1.9 < z < 3.5$. Fig. \ref{fig:VIRUS} shows a rendering of VIRUS and its optical design. 

The VIRUS concept has already been proven by the Mitchell Spectrograph (formerly known as VIRUS-P; Ref. \citenum{Hill08b}), a single  prototype VIRUS spectrograph that has been in use at the McDonald Observatory 2.7 m Harlan J. Smith telescope since 2007. The instrument has excellent throughput in the blue down to 350 nm ($\sim30$\%, excluding the telescope and atmosphere). For VIRUS, similar or better throughput will be essential to keep the number of LAE detections sufficiently high for achieving the goals of HETDEX. At the bluest wavelengths, a limiting optical component is the VPH diffraction grating that is used as the instrument's dispersing element. Ref. \citenum{Adams08} discusses the performance of VPH gratings developed for the Mitchell Spectrograph, which at that time pushed the technology to the highest diffraction efficiency achieved at 350 nm ($\sim60$\%); for VIRUS, we desire even higher diffraction efficiency of $\sim70$\% at 350 nm (see $\S$\ref{sec:gratingspec}). In order to tune the VPH layer fabrication prescription for achieving such high efficiency and maintaining it across the instrument's spectral coverage, we must understand how a given diffraction grating distributes the incoming light into various diffraction orders and where losses (i.e., scattering and/or absorption) occur. In addition, a challenge that is currently specific to VIRUS is achieving consistency in the required high standard of performance for large batches of $>150$ gratings. 

In this paper, we discuss methods for testing and verifying the suite of $> 150$ VPH diffraction gratings for VIRUS. We begin in $\S$\ref{sec:gratingspec} by describing the production design of the gratings. In $\S$\ref{sec:performance}, we discuss the methods we have used to date in measuring the diffraction efficiency and highlight the measurements of select prototype gratings that have helped lead us toward an acceptable fabrication prescription. In $\S$\ref{sec:tester}, we present the design of a new test apparatus that will allow for the efficient verification of the diffraction efficiency specification for the entire suite of VIRUS gratings. Finally in $\S$\ref{sec:conclusions}, we outline the general conclusions of the paper.

\section{VIRUS VPH GRATING DESIGN} \label{sec:gratingspec}
In the last decade, VPH gratings have become the standard in astronomical spectroscopy as they provide higher diffraction efficiency and versatility, among other advantageous traits, over classic surface relief gratings\cite{Barden98}. For useful overviews of the physics of VPH gratings, we refer the reader to Refs. \citenum{Arns99}, \citenum{Barden00}, and \citenum{Baldry04}. Ref. \citenum{Barden98} gives a brief overview of how VPH gratings are fabricated. 

Thanks to the extensive testing of VPH gratings at our test facility\cite{Adams08} and that the Mitchell Spectrograph allows for tuning of the collimator angle\cite{Hill08b}, different realizations of the VIRUS grating design have effectively been tested in the field. With the additional test results we present in the following section, we have arrived at a suitable design for the production VIRUS grating. As mentioned above, the key properties of the gratings are high diffraction efficiency (especially for the bluer wavelengths) and repeatability of the grating properties from unit to unit. These traits are essential for the ability of HETDEX to achieve its goals. 

\begin{table}[t!]
\caption{VIRUS external diffraction efficiency requirement for unpolarized light.} 
\label{tab:Efficiency}
\begin{center}       
\begin{tabular}{|c|c|c|} 
\hline
\rule[-1ex]{0pt}{3.5ex}  \textbf{$\lambda$ (nm)} & \textbf{Batch Mean} & \textbf{Batch Minimum} \\
\hline
\rule[-1ex]{0pt}{3.5ex}  350 & 70\% & 60\% \\
\hline
\rule[-1ex]{0pt}{3.5ex}  450 & 70\% & 60\% \\
\hline
\rule[-1ex]{0pt}{3.5ex}  550 & 40\% & 30\% \\
\hline
\end{tabular}
\end{center}
\end{table}
   \begin{figure}[t]
   \begin{center}
   \begin{tabular}{c}
   \includegraphics[width=0.95\textwidth]{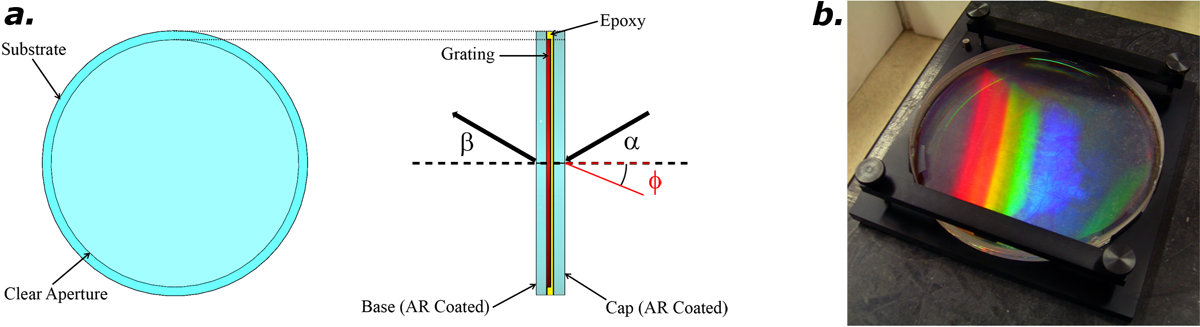}
   \end{tabular}
   \end{center}
   \caption[example] 
   { \label{fig:GratingSchem} 
\textit{a}) A schematic drawing of the VIRUS VPH diffraction grating. At left is a face-on view of the grating and at right is an edge-on view. For a sense of scale, the substrate diameter is 148 mm and the total thickness of the grating assembly is 16 mm. Note that the VPH layer thickness is exaggerated and is not to scale. The incident angle $\alpha$ and angle of diffraction $\beta$ are shown in addition to the direction of the fringe tilt, indicated in red by $\phi$. \textit{b}) A photograph of a prototype VIRUS VPH grating constructed to the specifications in $\S$\ref{sec:gratingspec}. The grating is mounted in a cell for testing in the apparatus that is discussed in $\S$\ref{sec:tester}.
}
   \end{figure} 

A schematic drawing of a VIRUS production grating can be seen in Fig. \ref{fig:GratingSchem}$a$. The grating assembly has physical dimensions of 148 mm in diameter $\times$ 16 mm in total thickness; the VPH layer itself has a 138 mm diameter clear aperture and is sandwiched between two 8 mm thick, anti-reflection (AR) coated fused silica substrates using an optical grade adhesive. The grating has a fringe frequency of 930 lines mm$^{-1}$ and operates at order $m=1$ in transmission from 350 to 550 nm for unpolarized light. Due to the large number units required for VIRUS ($\sim200$ gratings maximum when accounting for spares and witness samples), the gratings will be delivered from the vendor in batch sizes of up to $\sim70$ units, but no smaller than 10. The external diffraction efficiency of the entire grating assembly for unpolarized light is thus defined as a mean over a given batch. In addition there is a minimum allowable efficiency for any given grating in a batch. The batch mean and minimum external diffraction efficiencies are shown in Table \ref{tab:Efficiency}. Due to imperfections in the fabrication process, it is possible to have spatially variable efficiency across the grating clear aperture, the amount of which can be dependent on the size of the sub-aperture used for testing. As such, the diffraction efficiency specification in Table \ref{tab:Efficiency} is given as the average of multiple sub-apertures spread that are over the grating's clear aperture (see $\S$\ref{sec:tester} for how this specification will be verified). 

For a peak diffraction efficiency between 350 and 400 nm, it can be easily shown through the Bragg condition (e.g., see Ref. \citenum{Baldry04}) given the parameters above that the grating angle of incidence $\alpha$ should optimally be $\sim10$\degree. However, Ref. \citenum{Burgh07} has shown that the location in detector space where the wavelength satisfying the Bragg condition is imaged is also the location of the ``Littrow recombination ghost''. This ghost is caused by reflection off the CCD, recollimation by the camera, followed by either 1) $m=1$ reflective recombination, or 2) $m=1$ transmissive recombination followed by subsequent reflection off the inner surface of the front grating substrate (i.e., the cap in Fig. \ref{fig:GratingSchem}$a$) and finally $m=0$ transmission. Following either case 1 or 2, the recombined light makes a final pass through the camera and is finally reimaged on the detector. While the monochromatic efficiency of the ghost is quite low (on the order of $\lesssim0.1$\%), its wavelength integrated strength can dominate the signal in a given resolution element of the direct spectrum. As discussed in Ref. \citenum{Adams08}, this ghost can masquerade as a solitary emission line source, which for HETDEX could contribute significantly to sample contamination since normal star-forming LAE detections in our redshift range do not include emission lines other than \lya. Although the location of the ghost can easily be predicted in detector space for a bright source in a given fiber, we choose to avoid the issue all together by introducing a tilt $\phi = -1$\degree\ to the fringes, as suggested in Ref. \citenum{Burgh07}. This effectively decouples the Bragg condition from the Littrow configuration. Note that we have adopted the sign convention of Ref. \citenum{Burgh07} for $\phi$, where negative tilts move the plane of the fringes away from the incident beam, thereby effectively increasing the angle of incidence on the grating layer. The fringe tilt direction with respect to the incoming light is depicted in Fig. \ref{fig:GratingSchem}$a$. When including $\phi$, we can reduce the angle of incidence on the grating substrate to 9\degree. This roughly provides for the retention of the diffraction efficiency curve of a grating with unslanted fringes and $\alpha=10$\degree\ while pushing the Littrow ghost off the CCD detector as a result of the change in the physical grating angle. 

\section{EVALUATING GRATING PERFORMANCE} \label{sec:performance}

\subsection{Prototype VIRUS Production Gratings}\label{subsec:prototypegratings}
   \begin{figure}[t]
   \begin{center}
   \begin{tabular}{c}
   \includegraphics[width=\textwidth]{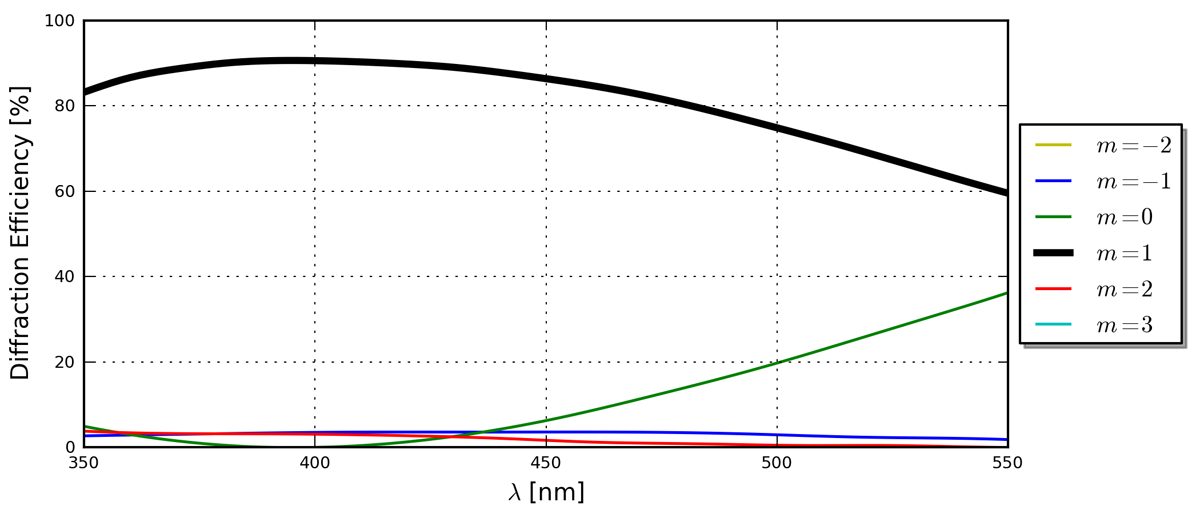}
   \end{tabular}
   \end{center}
   \caption[example] 
   { \label{fig:RCWA} 
The theoretical performance of the VIRUS prototype VPH diffraction gratings calculated using RCWA for $-2\leq m \leq 3$. The curves are for a grating with the specifications described in $\S$\ref{sec:gratingspec} and $\S$\ref{subsec:prototypegratings} (i.e., 930 lines mm$^{-1}$, $\alpha = 9$\degree, $\phi = -1$\degree, $d = 5.5$ $\mu$m, and $\Delta n = 0.037$; we assume that $n_{\mathrm{DCG}} = 1.45$). For a fair comparison with measurements, the modeled efficiency curves are shown after being multiplied by the measured transmission of the AR coated substrates and epoxy ($T_{\mathrm{sub}}(\lambda)$) as well as for the transmission of a typical DCG layer ($T_{\mathrm{DCG}}(\lambda)$; see Fig. \ref{fig:firstorder} and $\S$\ref{subsec:1storder}).
}
   \end{figure} 
To carry out a design study and to produce diffraction gratings for the initial tests\cite{Tuttle12} of the VIRUS spectrographs before entering the mass production phase, we have fabricated four VPH gratings according to the specification described in $\S$\ref{sec:gratingspec}. To achieve a sufficiently wide efficiency bandwidth (e.g., see Ref. \citenum{Barden00} in the Kogelnik approximation\cite{Kogelnik69}) for meeting the diffraction efficiency requirements outlined in the previous section, the \textit{targeted} parameter values for the dichromated gelatin (DCG) diffracting layer are optical thickness $d = 5.5$ $\mu$m and index of refraction modulation $\Delta n = 0.037$ (which is assumed to be sinusoidal). In Fig. \ref{fig:RCWA}, we show the theoretical diffraction efficiency of such a grating for $-2 \leq m \leq 3$, which will be useful for later comparison with measurements. These efficiency curves were calculated using Rigorous Coupled Wave Analysis (RCWA\cite{Gaylord85}) assuming an average index of refraction of the diffracting layer of $n_{\mathrm{DCG}} = 1.45$. 

The four gratings are referred to hereafter by their serial numbers (327-3, 338-1, 412-2, and 412-4) which identify them by procurement batch and allows them to be traced to a specific fabrication process. For verification of the AR coating and epoxy transmission, we have also fabricated a blank set of AR coated substrates (that contain no DCG) which have been bonded together using the same optical epoxy as the gratings. A photo of one of the prototype VIRUS gratings can be seen in Fig. \ref{fig:GratingSchem}$b$.

\subsection{The Grating Test Bench} \label{subsec:testbench}
To test the prototype gratings, we utilize an automated test facility that was first introduced and developed in Ref. \citenum{Adams08}. We refer the reader to that work for the details of its design and operating procedures. In short, the design is very similar to the tunable spectrograph concept that is shown in Ref. \citenum{Barden98}, Fig. 10. A commercial monochromator with $R \approx 240$ at 500 nm is used containing a 100 W quartz tungsten halogen (QTH) lamp as well as a 30 W deuterium lamp that are both constant in time at the $\lesssim1$\% level. The former, while bright and yielding excellent $S/N$ in our measurements, produces copious amounts of red light, some of which leaks through the monochromator especially biasing the bluer measurements ($\lambda \lesssim 450$ nm) that are intrinsically at a lower signal level\cite{Adams08}. The deuterium lamp is a clean source over the entire spectral range of interest, but is much less intense and causes problems with low $S/N$ when measuring diffraction orders where the grating's diffraction efficiency can be $\lesssim 10$\% (i.e., typically for $m \neq 1$). To avoid issues with low $S/N$ and leaking light at undesirable wavelengths, we use the QTH source in conjunction with a set of 10 nm FWHM narrow-band filters that are placed in the beam before the grating for up to six discrete wavelengths: 350, 370, 400, 450, 500, and 550 nm (at a minimum, we use 350, 450 and 550 nm for a given test). Additional modifications to the tester had to be made to accommodate the circular footprint of the VIRUS gratings (the gratings tested in Ref. \citenum{Adams08} for the Mitchell Spectrograph were rectangular with the fringes aligned with one of the substrate edges). This introduces an additional alignment step to the setup process for aligning the grating diffraction axis to the plane that is coincident with the motion of the photodiode detector. For this purpose, we have designed a custom cell that allows by-hand rotational adjustment of the grating about its normal, which can be verified by eye with a bright laser source for quick feedback. For the test results presented here, we utilize a 12.5 mm diameter collimated beam for consistency with the apparatus discussed in $\S$\ref{sec:tester}. Finally, we note that we have estimated the accuracy of the absolute zero-point alignment of the grating rotation stage (which sets $\alpha$) to be $\pm0.25$\degree.

\subsection{First Order Diffraction Efficiency}\label{subsec:1storder}
   \begin{figure}[t]
   \begin{center}
   \begin{tabular}{c}
   \includegraphics[width=\textwidth]{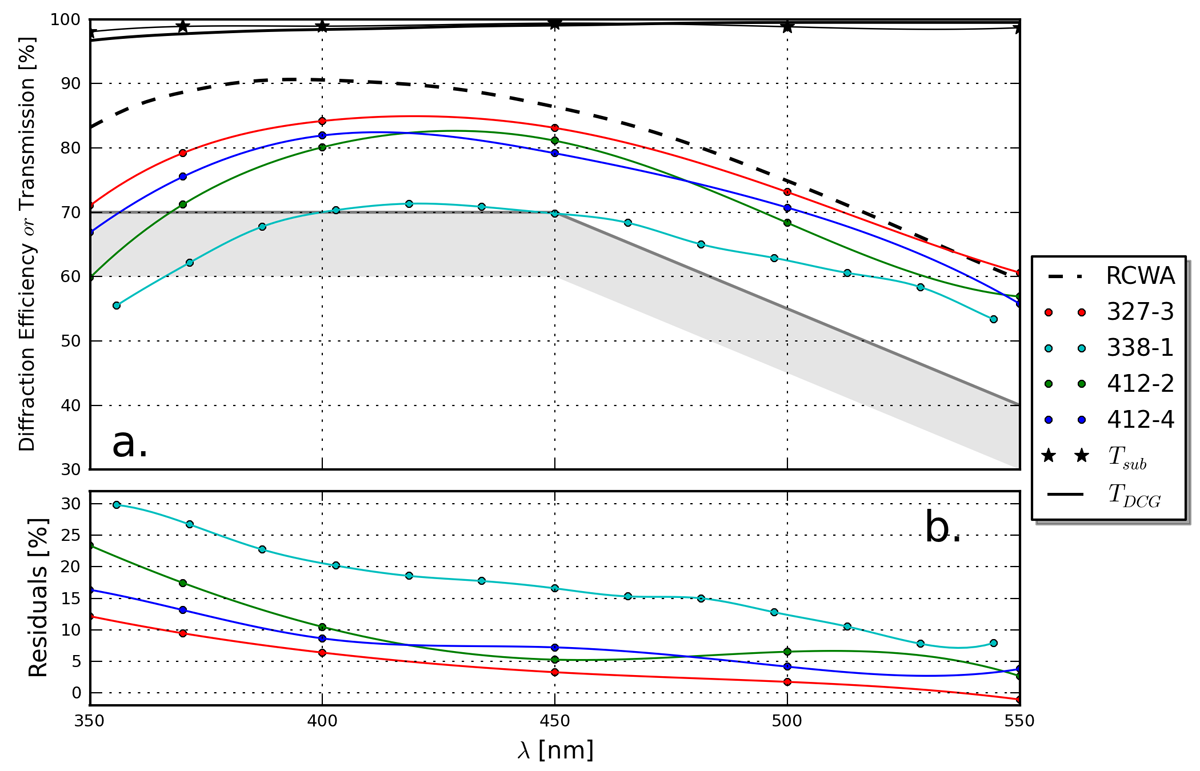}
   \end{tabular}
   \end{center}
   \caption[example] 
   { \label{fig:firstorder} 
\textit{a}) The measured first order external diffraction efficiency of the VIRUS prototype VPH diffraction gratings for $\alpha = 9$\degree\ with the fringes oriented as shown in Fig. \ref{fig:GratingSchem}$a$. We also show $T_{\mathrm{sub}}(\lambda)$ (the measured transmission of the AR coated substrates and epoxy) and $T_{\mathrm{DCG}}(\lambda)$ (the transmission of a typical DCG layer; see $\S$\ref{subsec:1storder}). The grey line shows the batch mean external diffraction efficiency requirement for the VIRUS gratings while the lower bound of the grey shaded region indicates the batch minimum (see Table \ref{tab:Efficiency}). \textit{b}) The residuals between the RCWA calculated $m=1$ efficiency curve from Fig. \ref{fig:RCWA} and the data for each grating. For both panels, the solid curves connecting the data points are cubic spline fits to the data. 
}
   \end{figure} 
For each of the prototype VIRUS production gratings listed in $\S$\ref{subsec:prototypegratings}, we have measured the first order external diffraction efficiency for a single 12.5 mm diameter beam at $\alpha=9$\degree\ with the grating test bench. For each of these tests, we have oriented the grating in the beam such that the tilted fringes match the orientation shown in Fig. \ref{fig:GratingSchem}$a$. Each grating was measured using the narrow-band filtered QTH source, except for 338-1 which was measured with the deuterium source before we obtained the narrow-band filters. The results are shown in Fig. \ref{fig:firstorder}$a$. As can be seen, all four gratings meet or exceed the batch mean external diffraction efficiency requirement at 450 and 550 nm. However, the performance appears to fall off towards 350 nm where only one grating (327-3) exceeds the batch mean requirement. While two additional gratings (412-2 and 412-4) would meet specification for the batch \textit{minimum} requirement at 350 nm, it appears that the mean of our small sample of 4 gratings would not meet the batch \textit{mean} requirement. Although the fabrication prescription is the same for all four gratings, the performance of 338-1 is significantly worse for all wavelengths as compared to the other gratings. This is due to the final processing of the grating layer, in which this particular grating was exposed to heat during the drying procedure for an extended period of time. This caused the optical thickness of the diffracting layer to become too thin by $\sim2$ $\mu$m. Thus, we generally exclude 338-1 from what we consider the result of a typical grating made with the intended prescription. 

As seen in Fig. \ref{fig:firstorder}$a$, we have measured the transmission of the AR coated and epoxied blank substrates $T_{\mathrm{sub}}(\lambda)$ at 9\degree\ angle of incidence. We find an average transmission of 97.9\%, which does not vary significantly with wavelength. We remind the reader that the blank epoxied substrates do not contain any DCG, so the latter measurements do not take into account the transmittance of the grating layer. At the time of our tests, we did not have our own measure of the DCG transmission of our gratings. We thus estimated it by utilizing the transmission data presented in Fig. 2 of Ref. \citenum{Barden00}, which is for a uniformly exposed 15 $\mu$m thick layer. We measured the physical thickness of the DCG layer in our gratings to be 7.5 $\mu$m and thus estimate their DCG transmission $T_{\mathrm{DCG}}(\lambda)$ to be that from Ref. \citenum{Barden00} taken to the power of 7.5 $\mu$m/15 $\mu$m $= 0.5$. The result is plotted in Fig. \ref{fig:firstorder}$a$. For a fair comparison with the data, all RCWA curves shown in this work have been multiplied by $T_{\mathrm{sub}}(\lambda)$ and $T_{\mathrm{DCG}}(\lambda)$. For all four gratings, we see a general trend of increasing deviation from the RCWA predicted efficiency (after correction for the substrate, epoxy, and DCG transmittance) with decreasing wavelength, especially for $\lambda \lesssim 400$ nm (see Fig. \ref{fig:firstorder}$b$). 

Given these $m=1$ results, we have determined that fabricating a grating that can meet the requirements outlined in $\S$\ref{sec:gratingspec} is possible for our prescription and process (e.g., see 327-3). In addition, the fabrication process is consistent enough to produce gratings that are similar to each other within $\sim5$\% for most of the spectral coverage of the instrument. However, the decreased performance of all gratings towards 350 nm (which is systematic relative to the RCWA predictions) and that only one of the gratings performs better than our batch mean requirement at that wavelength beckons further investigation. Before continuing, we mention that since we have only measured a single sub-aperture that is only $\sim10$\% the diameter of the grating's clear aperture, we cannot rule out that spatial variability of the diffraction efficiency may be a cause of the apparent decreased performance at 350 nm. Unfortunately, the grating test bench does not easily allow for a useful number of sub-apertures distributed evenly across the clear aperture to efficiently be measured for a better estimation of the grating's characteristic diffraction efficiency without introducing significant systematic errors. See $\S$\ref{subsec:performance} for further discussion on this issue.   

\subsection{Locating All of the Incident Light}\label{subsec:ordertests}
To explore the discrepancy between the $m=1$ RCWA prediction and the $m=1$ data for all four gratings, we look to additional spectral orders in an attempt to account for all of the incident light. The test bench is a versatile and general-use test facility that can measure the diffraction efficiency as a function of $\lambda$, $\alpha$, and $m$. For our prototype VIRUS gratings at $\alpha = 9$\degree\ and $350 < \lambda \mathrm{(nm)} < 550$, we can generally access $-2 \leq m \leq 3$ with the tester when a solution for the grating equation exists and the diffraction angle $\beta$ is within $\pm65$\degree. For our six possible discrete wavelengths, this translates into $-2 \leq m \leq 3$ for 350 and 370 nm (note that this range covers all defined transmissive spectral orders at $\alpha=9$\degree\ for these wavelengths) and $-1 \leq m \leq 2$ for 400, 450, 500, and 550 nm. 

   \begin{figure}[t]
   \begin{center}
   \begin{tabular}{c}
   \includegraphics[width=\textwidth]{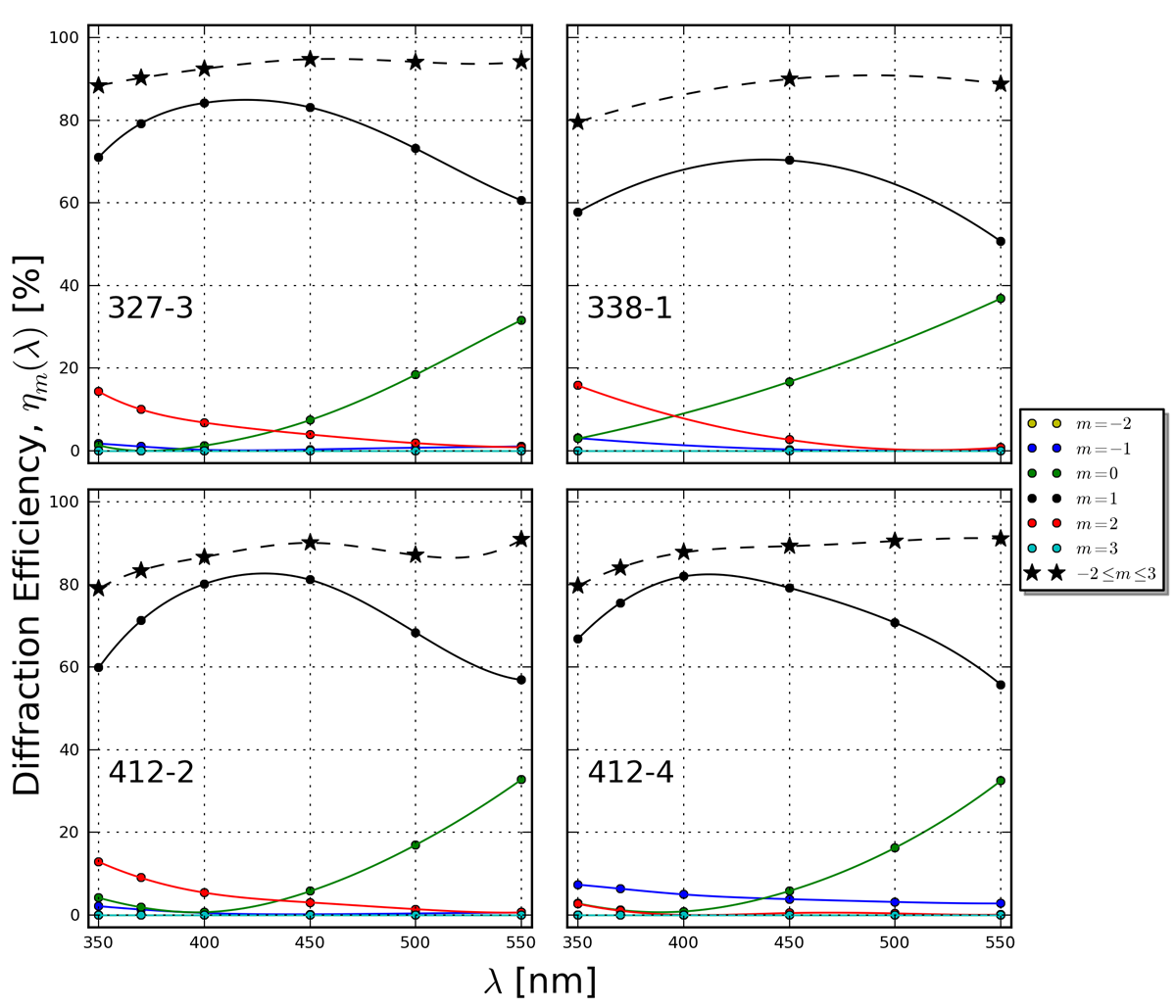}
   \end{tabular}
   \end{center}
   \caption[example] 
   { \label{fig:multiorder} 
The external diffraction efficiency of all four gratings measured for $-2 \leq m \leq 3$ at $\alpha=9$\degree\ with the fringes oriented as shown in Fig. \ref{fig:GratingSchem}$a$. For easy comparison with the RCWA predictions, the color scheme is the same as for Fig. \ref{fig:RCWA}. The curves connecting the data points are spline fits to the data (cubic, except for 338-1, which is quadratic due to having measured only three wavelengths). Note that the black stars represent the sum of all light found in all measured diffraction orders. 
}
   \end{figure} 

In Fig. \ref{fig:multiorder}, we show the $\alpha=9$\degree\ external diffraction efficiency measurements for our four prototype gratings over $-2 \leq m \leq 3$. The measurements for $m\neq1$ were made for the same 12.5 mm diameter sub-aperture as the $m=1$ data presented in the previous section. For 338-1, we have remeasured the $m=1$ efficiency using the narrow-band filtered QTH source, which agrees with the deuterium-based measurements shown in Fig. \ref{fig:firstorder} within $\pm2$\% for the measured wavelengths. We first note that each grating's multi-order diffraction efficiency plot appears very similar to the RCWA predictions shown in Fig. \ref{fig:RCWA}. For example, we have measured no signal for any grating in $m=-2$ or $3$, which is consistent with the RCWA predictions. Additionally, the main thief from the $m=1$ efficiency is $m=0$, especially for redder wavelengths. For bluer wavelengths, increased $m=2$ efficiency is often the dominant order that removes light from $m=1$. Additionally, a small but non-negligible amount of light is often diffracted into $m=-1$. While the general trends from the RCWA models can be observed in our data, the main discrepancy lies in the absolute efficiency at which the diffraction occurs. This is most strikingly true for $m=2$ for 327-3, 338-1, and 412-2, which all display efficiencies in excess of the $m=2$ RCWA predictions by as much as $10$\%. Additionally, 412-4 appears to be an outlier by having relatively low $m=2$ efficiency (which closely matches the RCWA predictions) and a relatively high $m=-1$ efficiency. This is extremely peculiar as 412-2 and 412-4 were fabricated by the same prescription and process. We note that while we can qualitatively reproduce these $m=-1$ and $2$ efficiency curves individually with a RCWA model different from that shown in Fig. \ref{fig:RCWA}, we cannot do so without severely affecting the $m=1$ curve such that it becomes totally mismatched from the data. While detailing the discrepancies between the models and the data warrants further investigation, it is beyond the scope of this paper.

\begin{figure}[t]
   \begin{center}
   \begin{tabular}{c}
   \includegraphics[width=\textwidth]{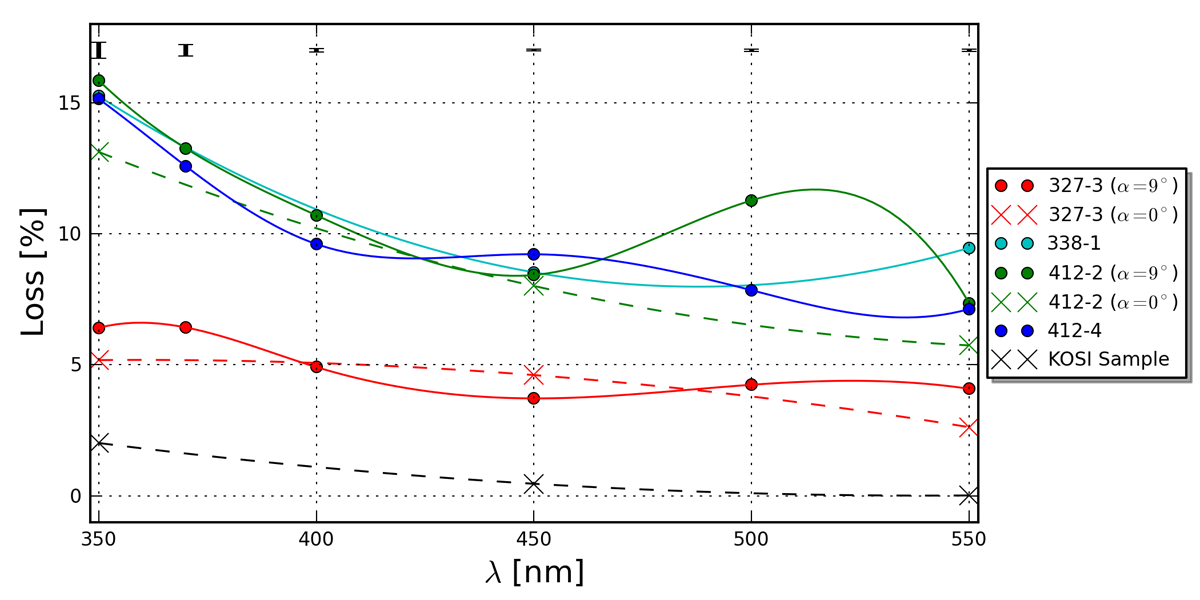}
   \end{tabular}
   \end{center}
   \caption[example] 
   { \label{fig:loss} 
The diffractive loss, as calculated from Eq. \ref{eq:loss} for all four prototype gratings and the sample KOSI VPH grating. The filled circles represent the loss as measured with the grating at $\alpha=9$\degree\ while the ``X''s represent measurements at $\alpha=0$\degree. The curves connecting the data points are spline fits to the data (cubic, or quadratic if the number of data points is three). Above each discrete wavelength, the black error bar indicates the measured level of photometric error as formally propagated through Eq. \ref{eq:loss}. For wavelengths where the $S/N$ is high (i.e., $\lambda \geq 400$ nm), the photometric errors are likely negligible compared to systematic errors caused by the uncertainty in $\alpha$ (see $\S$\ref{subsec:testbench}). 
}
   \end{figure}

In Fig. \ref{fig:multiorder} for all gratings, we have also plotted the \textit{total} measured external diffraction efficiency over $-2 \leq m \leq 3$ (which for all intensive purposes corresponds to all possible transmissive diffraction orders). As can be seen, the total transmissive diffraction efficiency displays a ubiquitous drop towards 350 nm while showing asymptotic behavior with increasing wavelength. Since the total diffraction efficiency of a grating at any wavelength can be no greater than the product $T_{\mathrm{sub}}(\lambda) T_{\mathrm{DCG}}(\lambda)$ (the value of which is $>94$\% for all wavelengths based on our measurements and assumptions about the DCG layer), the data imply that some of the incident light is still unaccounted for. To quantify this ``diffractive loss'', we calculate the following:
\begin{equation}\label{eq:loss}
\mathrm{Loss}(\lambda) = T_{\mathrm{sub}}(\lambda)\: T_{\mathrm{DCG}}(\lambda)\: - \displaystyle\sum_{m = -2}^{3} \eta_{m}(\lambda) \; ,
\end{equation}
where $\eta_{m}$ is the external diffraction efficiency measured for transmissive order $m$. In Fig. \ref{fig:loss}, we show the diffractive loss for all four gratings. As indicated by the total diffraction efficiency curves in Fig. \ref{fig:multiorder}, the loss is ubiquitously greatest at the shortest wavelengths and appears to level as one approaches 550 nm. Additionally, there is a striking difference between 327-3 and the other three gratings. 327-3 has a much flatter loss curve that is systematically lower at all wavelengths as compared to the other three gratings. For 327-3, we can account for all but $\sim5$\% of the incident light in the transmissive diffraction orders. The losses, as measured for the remaining three gratings, are all consistent with each other and reach as high as $\sim15$\%. The systematic difference between 327-3 and the other three gratings is quite interesting, especially since 327-3, 412-2, and 412-4 display very similar performance for $m=1$. Similarly, that the loss properties of 338-1 are very similar to 412-2 and 412-4 is intriguing since 338-1 is an outlier in terms of its relatively poor $m=1$ performance. 

Our calculations have taken into account the transmission of the AR coated fused silica substrates, optical epoxy, and the DCG layer as well as the light diffracted into all transmissive orders. We envisage two possibilities for explaining the diffractive loss and the differences/similarities in this quantity between the gratings. First, as the only quantity we have thus far not measured directly when calculating the loss, it is possible that $T_{\mathrm{DCG}}(\lambda)$ from Ref. \citenum{Barden00} is incorrect for our gratings due to some different processing technique or unknown contaminant introduced to the grating layer during fabrication. Second, it is possible that the VPH gratings have small but non-negligible total diffraction efficiency for reflective orders. 

We can infer that reflective diffraction is relatively inefficient for our gratings given that we have accounted for all but $\lesssim15$\% of the incident light in the transmissive diffraction orders and the transmission of various other optical components. While we cannot directly measure the reflective orders with the test bench due to the limitations of its range of motion, we can at least put constraints on it by utlizing the Littrow recombination ghost\cite{Burgh07}. As discussed in $\S$\ref{sec:gratingspec} and in Ref. \citenum{Burgh07}, the Littrow ghost can occur for two modes of recombination: reflective and transmissive. For VPH gratings with a wedged cap substrate (i.e., a grism), the reflective and transmissive recombination ghosts can be spatially decoupled at the focal plane. Since our gratings are immersed in plane parallel substrates, the locations of the ghosts produced by both recombination modes spatially coincide. While we therefore cannot measure the reflective recombination efficiency alone, we can measure the combined strength of the ghost produced by both modes and use that as an upper limit to determine if the diffractive losses of our gratings can be explained by unmeasured reflective diffraction orders. There are also additional caveats with this indirect method: 1) the reflective recombination ghost we measure is for the reflective $m=1$ only, and 2) since the ghost is recombined, we can only calculate the wavelength averaged reflective diffraction efficiency. As discussed in Ref. \citenum{Tuttle12}, we have begun to assemble VIRUS unit spectrographs for testing before beginning mass production. By design, the $\alpha=9$\degree\ setup does not image the Littrow ghost. As such, we have set up a single VIRUS spectrograph containing a prototype grating at $\alpha = 10$\degree, which images the Littrow ghost near the edge of the CCD chip (since this is close to the design angle of incidence, we do not expect the results to be significantly different between the two grating setups). Following Ref. \citenum{Burgh07}, we have measured the total flux in the direct spectrum from a single fiber illuminated by a QTH source as well as the flux in the corresponding Littrow ghost. The ratio of the ghost flux to the integrated flux in the direct spectrum is $2.0\times10^{-5}$. Assuming no absorption in the CCD substrate, the CCD reflectivity averaged over $350 < \lambda (\mathrm{nm}) < 550$ is 12\% and we estimate the camera throughput to be 67\% (which includes the effect of the camera's central obstruction and the transmissivity/reflectivity of its optics). This results in a wavelength averaged grating recombination efficiency of 0.04\%, which should be typical for all gratings presented in this work. Even if the recombination efficiency was due solely to the reflection mode, we can conclude from this that the $m=1$ reflective diffraction efficiency contributes negligibly to making up the difference measured in the loss. While other reflective orders remain unconstrained, it is likely that any case of reflective diffraction will be similarly inefficient due to being so far away from the Bragg condition for the orientation of the grating's fringes.

Given the above evidence for very small reflective diffraction efficiency, we turn again to the DCG layer. As stated above, it is possible that the measurements of $T_{\mathrm{DCG}}(\lambda)$ from Ref. \citenum{Barden00} are not appropriate for our gratings. Differences in $T_{\mathrm{DCG}}(\lambda)$ could also explain the systematic offset of the 327-3 loss from that of the other three gratings. Since 327-3 was fabricated in an earlier batch, it is possible that something changed in the DCG processing to cause the grating layer to become less transparent (e.g, increased scatter or absorption) for the later gratings. Since we were not originally supplied with any processed uniformly exposed DCG for measuring $T_{\mathrm{DCG}}(\lambda)$ directly for each grating batch, we cannot retroactively measure with high certainty. As additional evidence that suggests that the use of an incorrect DCG layer transmission may be a cause of the loss, we have measured the diffractive loss for various $\alpha$ and found that it remains constant. In Fig. \ref{fig:loss}, we show the loss as measured for $\alpha=0$\degree\ for 327-3 and 412-2. There are small differences between the $\alpha=0$\degree\ and $\alpha=9$\degree\ data, but they are on the level that might be expected from the $\pm0.25$\degree uncertainty in $\alpha$ for a given test (this uncertainty is largest away from the Bragg wavelength at $\sim400$ nm). The independence of the measured loss with $\alpha$ suggests some common mode of loss that is independent of the physical diffraction process, which points towards the DCG layer transmission since we have not explicitly measured it for our gratings. As additional evidence, we have measured the loss for an additional sample grating that was provided to us by Kaiser Optical Systems, Inc. (KOSI) during the grating development phase for the Mitchell Spectrograph. This grating is a small 25 mm $\times$ 20 mm sample with 831 lines mm$^{-1}$. We do not know of the details of the grating layer other than that the grating is optimized for use at $\alpha=10$\degree and has unslanted fringes. Also sandwiched in the same uncoated fused silica 76 mm $\times$ 25 mm rectangular substrate is a sample of uniformly exposed DCG that was processed simultaneously and identically to the grating. Measurement of the uniformly exposed DCG sample gives $T_{\mathrm{sub}}(\lambda) T_{\mathrm{DCG}}(\lambda)$ directly and we have measured all available transmissive diffraction orders at $\alpha=0$\degree. We have thus calculated the diffractive loss for this grating using these data and have plotted the result in Fig. \ref{fig:loss}. As can be seen, we are able to account for nearly all the incident light for this grating, with a maximum of 2.0\% loss at 350 nm. The wavelength averaged loss is 0.83\%, which is closer to the order of magnitude one might expect for the total reflective diffraction efficiency. 

\subsection{Meeting the Diffraction Efficiency Requirement}\label{subsec:meetspec}
While the task of accounting for all of the incident light on the grating was largely academic, it will prove useful in further iterations on the grating design. If in fact the higher diffractive loss in our four gratings is due to a decreased grating layer transmission as compared to that measured in Ref. \citenum{Barden00} after scaling for the DCG thickness, tracking down the cause could prove highly beneficial for increasing the diffraction efficiency. For example, 412-2 and 412-4 have similar $m=1$ external diffraction efficiency to 327-3 despite having $\sim5-10$\% higher measured loss. If $T_{\mathrm{DCG}}(\lambda)$ for these two gratings could be increased to the level as indicated by the 327-3 loss curve, each of these two gratings would at least meet the VIRUS batch mean external diffraction efficiency requirement at 350 nm where they currently fail to do so. Determining if this is the case will require the fabrication of a uniformly exposed DCG sample with identical post-processing for measurement. 

Just prior to the submission of this manuscript, we were able to receive a new uniformly exposed sample of DCG that is the same physical thickness as our gratings and made on the same process. Our transmission measurements (which were corrected for the glass substrate on which the DCG was processed) qualitatively match the loss shown in Fig. \ref{fig:loss} for gratings 338-1, 412-2 and 412-4 with transmission at 350, 450, and 550 nm of around 87, 92, and 95\%, respectively. Because these measurements were made with haste, we do not have a good sense of our errors and we refrain from quoting any further numerical transmission results. By adding additional processing steps to harden the gelatin in additional samples, the transmission appears to be systematically improving closer to $T_{\mathrm{DCG}}(\lambda)$ from Ref. \citenum{Barden00}. These results imply that the amount of absorption in the gelatin is normal and that the previous processing methods resulted in excessive scatter in the DCG film. We are currently working to further improve the DCG film scatter before proceeding with the production of VIRUS gratings. Note that with an improved process that approaches $T_{\mathrm{DCG}}(\lambda)$ from Ref. \citenum{Barden00}, we estimate that our latest gratings (based on the average properties of 412-2 and 412-4) would have $m=1$ efficiencies of about 73, 87, and 60\% at 350, 450, and 550 nm, respectively. This should exceed our requirements at all wavelengths and prove extremely beneficial for HETDEX science. 

   \begin{figure}[t]
   \begin{center}
   \begin{tabular}{c}
   \includegraphics[width=0.8\textwidth]{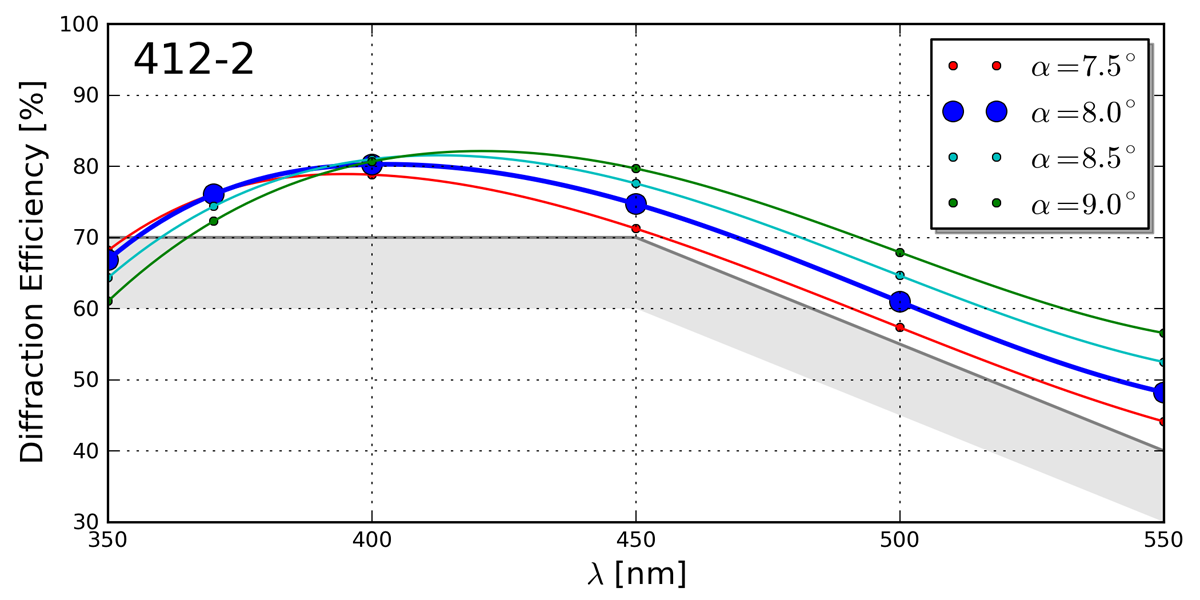}
   \end{tabular}
   \end{center}
   \caption[example] 
   { \label{fig:NoFT} 
The measured external diffraction efficiency of 412-2 for four different values of $\alpha \leq 9$\degree. These measurements show that the 350 nm external diffraction efficiency requirement can be more closely met with the same fabrication prescription as the prototype gratings and without resolving the DCG layer transmission issues by effectively reducing the angle of incidence. A simple way of achieving this would be to remove the fringe tilt ($\phi = 0$\degree) and keep $\alpha = 9$\degree. This scenario is closely represented by the thick blue efficiency curve, which is for 412-2 (i.e., a grating with $\phi = -1$\degree) oriented at $\alpha = 8$\degree. The solid curves connecting the data points are cubic spline fits to the data. 
}
   \end{figure} 

While it seems that we have identified the major culprit for the increased loss in our gratings (especially for the bluer wavelengths), the simple fact is that the prototype gratings nearly perform at the required level for the mode in which they will be used for VIRUS despite the increased scatter in the DCG layer. As seen in Fig. \ref{fig:firstorder}, all four gratings have a peak external diffraction efficiency that is slightly too red as compared to the desired efficiency curve represented by the RCWA model. We have seen that this is due in part to the DCG layer transmission (i.e., increased scatter) and that the grating diffracts an excess of light for bluer wavelengths into mainly $m=2$ ($m=-1$ for 412-4); the latter appears to be at odds with the RCWA models. Naturally through the Bragg condition, it can be seen that changing the angle of incidence on the grating can also shift the wavelength of peak efficiency (i.e., the Bragg wavelength), which can be accomplished by changing either $\alpha$ or $\phi$. In fact, a review of the setup of 327-3 during its exposure to the interference pattern during fabrication shows that it actually has $\phi$ closer to $-0.6$\degree. With its inferred higher DCG transmission, this helps explain why it has a slightly higher $m=1$ diffraction efficiency at 350 nm as compared to 412-2 and 412-4 (which both were setup properly to give $\phi = -1$\degree). In this way, it is possible to meet the VIRUS batch mean external diffraction efficiency requirement with the current grating prescription and fabrication techniques by simply setting $\phi = 0$\degree\ and keeping $\alpha = 9$\degree. Such a setup would boost the $m=1$ efficiency at 350 nm at the expense of the efficiency at 450 and 550 nm. This is acceptable since both of these wavelengths currently are exceeding the requirement by $\gtrsim10$\%. To see the magnitude of such a change, the $\phi=0$\degree and $\alpha=9$\degree\ scenario can be approximately represented by using our $\phi=-1$\degree\ gratings at $\alpha=8$\degree. As seen in Fig. \ref{fig:NoFT} for 412-2, this results in a boost in the 350 nm efficiency of $\sim6$\% while maintaining greater-than-required efficiency at higher wavelengths. The unslanted fringe design is also beneficial because the grating is simpler to fabricate and it makes the installation of the grating in the VIRUS spectrogaphs much more straightforward. Resolving the final DCG layer processing issues and/or deciding whether to remove the fringe tilt will be the final iterations required before proceeding with the grating production. 

\section{THE VIRUS GRATING TESTER} \label{sec:tester}

\subsection{Acceptance Test Methodology} \label{subsec:methodology}
The results presented above for the VIRUS prototype gratings constitute a full characterization of their diffraction efficiency performance and have helped provide essential feedback into the design and fabrication of the production gratings. Given the results at the designed $\alpha$ and $m$ for 327-3, 412-2, and 412-4, it appears that we have found a grating prescription that will nearly meet our diffraction efficiency specification with consistency. The grating test bench\cite{Adams08} used for the tests in the previous section is an excellent general purpose tool for fully characterizing a grating's performance. However, such level of detail is not necessary for verifying the performance requirements of the entire suite of $\sim200$ production gratings, since we mainly are concerned with the external diffraction efficiency in the configuration used for VIRUS and its consistency across large batches of gratings. 

As part of the contract that will be awarded to a vendor for fabrication of the gratings, one of our responsibilities is to provide an on-site method of acceptance testing. Given the number of gratings in our order, such a test must first and foremost be time efficient (ideally taking $\lesssim10$ minutes per grating). Additionally, it must be stable and repeatable (i.e., the ideal apparatus should have no moving parts), the apparatus must be compact, and the test process must be simple to carry out for vendor personnel (i.e., the ideal test should have ``push-button'' operational simplicity). Building a copy of the test bench for the vendor was an option, but it typically takes $\gtrsim30$ minutes to test a grating for $m=1$ at a given $\alpha$. Most of this time is spent in setup and alignment of the various rotation stages and the grating. This alignment is a manual process that can introduce user error. Additionally, the number of moving components in the apparatus adds a level of complexity that is unnecessary for acceptance purposes. The test bench also does not provide an easy way of testing multiple sub-apertures of the grating for estimating the average diffraction efficiency across the clear aperture. Given these limitations, we opted to design a new test apparatus that is specialized for the verification of the VIRUS external diffraction efficiency specification. We present the design and first results from the prototype apparatus below. 

\subsection{Design} \label{subsec:design}
   \begin{figure}[t]
   \begin{center}
   \begin{tabular}{c}
   \includegraphics[width=\textwidth]{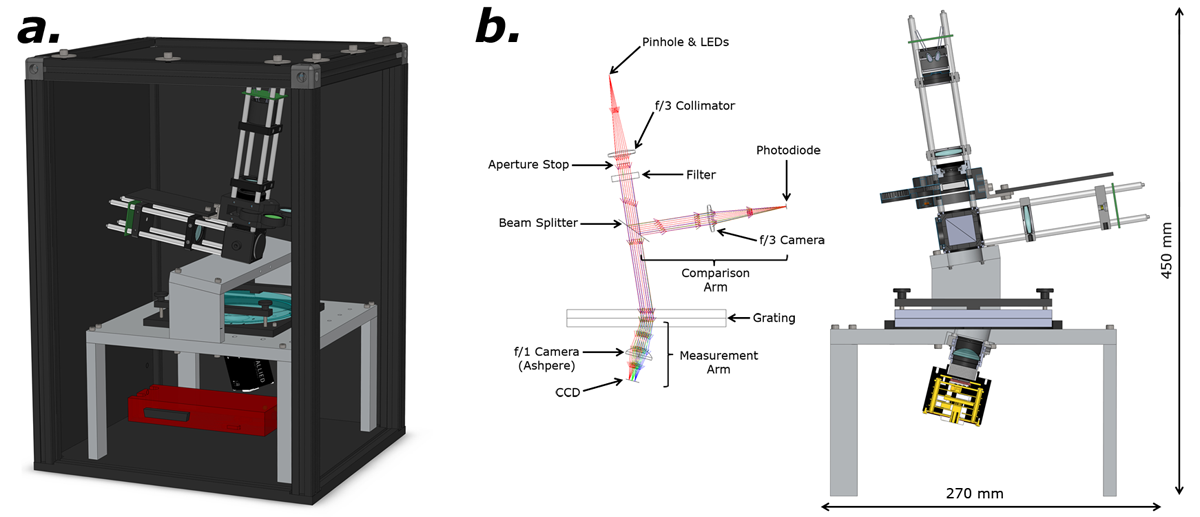}
   \end{tabular}
   \end{center}
   \caption[example] 
   { \label{fig:design} 
The opto-mechanical design of the VIRUS grating tester. $a$) A CAD rendering of the apparatus in its light-tight enclosure. The two open sides of the enclosure are closed during testing with double-layered blackout fabric. The red ``brick'' is a data acquisition unit, described in $\S$\ref{subsec:construction}. $b$) At left is a ray trace of the apparatus' optics with the major components labeled. To ease comparison between panel $a$ and the ray trace, a cross-section of the CAD model is shown at right with rough dimensions indicated for scale.
}
   \end{figure} 
Fig. \ref{fig:design} shows the opto-mechanical design of the VIRUS grating tester. Since our specification of the external diffraction efficiency is defined at three discrete wavelengths (see Table \ref{tab:Efficiency}), the apparatus will test only at those wavelengths utilizing LEDs as the light source. The LEDs are placed in an enclosure which includes an engineered diffuser before a 300 $\mu$m diameter pinhole that is placed at the focus of a stock\footnote{All stock lenses used were obtained from Edmund Optics and are UV-VIS coated fused silica.\\ Edmund Optics: \textit{http://www.edmundoptics.com}} 25 mm diameter $f$/3 singlet. The collimated beam is then stopped down with an adjustable aperture to 12.5 mm in diameter. While the LED light sources are centered on the wavelengths we wish to test, they are far from monochromatic, each having an emitted FWHM of approximately 25, 25, and 60 nm for the 350, 450 and 550 nm LEDs, respectively. We thus place a 10 nm FWHM narrow-band filter in the collimated beam for each respective LED. After the filter, the collimated beam is split 50:50 with a BK7 cube beam splitter. Half of the light is refocused with a stock 25 mm diameter $f$/3 singlet onto a silicon photodiode, which is used to self-calibrate for any instability in the LED output. The active area of the photodiode is 1.6 mm in diameter, which is large enough to accommodate the chromatic aberration attributed to the singlet lenses. The other half of the collimated beam is then incident on the diffraction grating at an angle $\alpha = 9$\degree. The diffracted light (which has an angle $\beta = 15.3$\degree\ at the 450 nm central wavelength) is then focused onto a 2/3''-format CCD by a stock 25 mm diameter $f$/1 aspheric lens. We choose the 12.5 mm beam diameter to eliminate any issue with differentially vignetting the dispersed collimated beam after the grating since the camera and collimator lenses have the same physical diameter. Photometry is performed on the CCD image of the refocused spot for each wavelength. The mechanical design of the tester fixes the collimator and camera angles. Thus, the flux measurements are calibrated through the measurement of a sample 930 line mm$^{-1}$ reference grating whose absolute external diffraction efficiency is known through careful measurement of the $m=1$, $\alpha = 9$\degree\ efficiency curve by the grating test bench facility\cite{Adams08}.

   \begin{figure}[t]
   \begin{center}
   \begin{tabular}{c}
   \includegraphics[width=\textwidth]{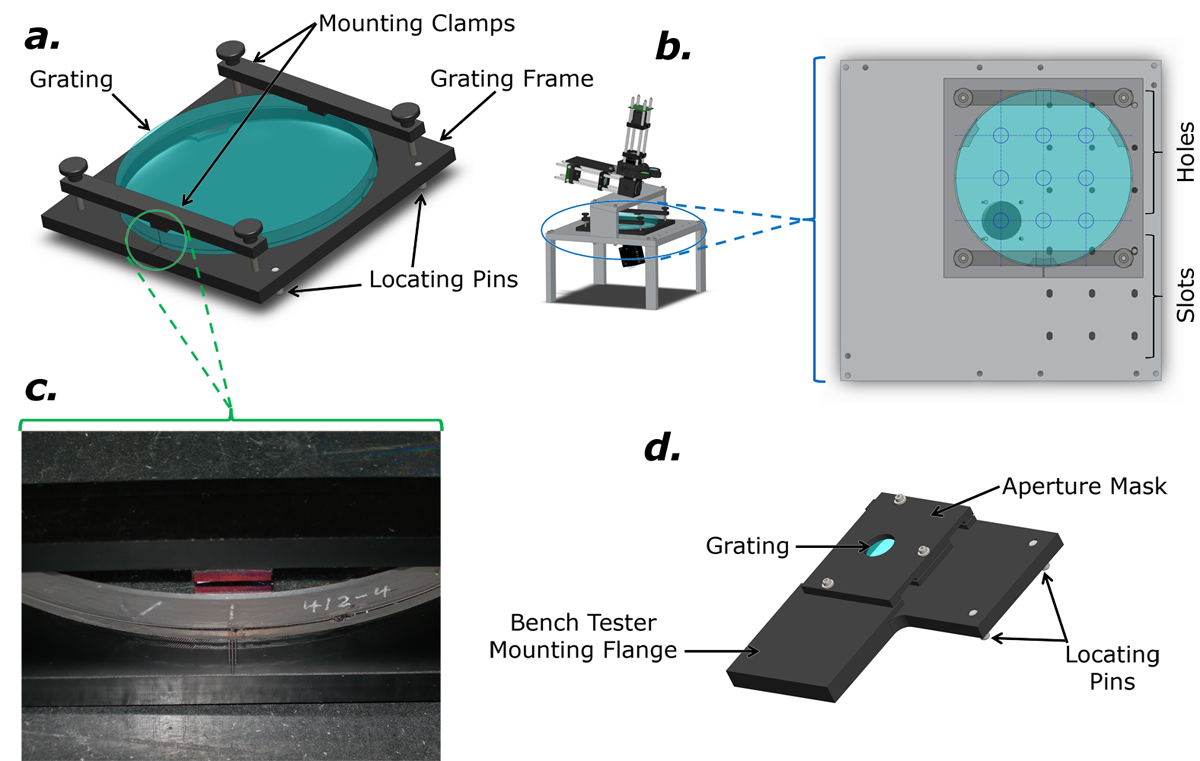}
   \end{tabular}
   \end{center}
   \caption[example] 
   { \label{fig:gratingcell} 
The mechanical design of the grating mounting cell. \textit{a}) A CAD rendering of the grating cell with major components labeled. \textit{b}) A CAD rendering of the tester base showing the series of holes and slots used for maintaining the rotational alignment of the grating when switching between the nine sub-apertures, which are indicated by the thin blue circles on the grating face. \textit{c}) A photo showing an engraving on the edge of one of the prototype VIRUS gratings that indicates the fringe orientation, including the direction of the tilt. This engraving is aligned with a matching feature on the grating frame for rapid initial alignment of the grating. \textit{d}) A CAD rendering of a mounting cell for a reference grating. This particular cell was designed for a square sample grating, and includes a built-in aperture mask to ensure that a common sub-aperture is measured on both of our testers for accurate calibration. As labeled, the cell design includes features for mounting on both of our testers. 
}
   \end{figure} 

We have attempted to simplify the opto-mechanical design of the tester by utilizing as many off-the-shelf components as possible. As such, most of the tester mechanics utilize the ThorLabs 30 mm Cage System\footnote{ThorLabs 30 mm Cage System: \textit{http://www.thorlabs.com/Navigation.cfm?Guide\_ID=2004}} and related hardware for mounting 1'' diameter optics. Where appropriate (e.g., for mounting the LEDs, the comparison photodiode, and their respective circuitry), we have modified stock cage components for our custom needs. As seen in Fig. \ref{fig:design}, we have designed custom aluminum hardware for fixing the collimator and camera angles to an accuracy of $\pm0.1$\degree\ that mates with the stock cage components. For testing multiple sub-apertures and to ease in the alignment of the diffraction grating, we have designed a delrin mounting cell in which the grating rests, detailed in Fig. \ref{fig:gratingcell}. The VIRUS grating specification requires the vendor to include a marking on the edge of the grating for determining the fringe tilt and fringe direction within $\pm1$\degree\ so that the grating can easily be placed in the tester in the correct orientation. A corresponding mark exists on the edge of mounting cell; when the markings are aligned by eye, the placement of the focused spot is guaranteed to be on the CCD chip for later fine adjustment. The cell also contains two press-fit locating pins which mate to a series of holes and slots on the tester base for constraining the rotational alignment of the grating cell as it is moved between a series of nine different test positions within the clear aperture (see Fig. \ref{fig:gratingcell}$b$). The tester is compact and can easily fit on a desk in an office environment. It has the approximate dimensions of 450 mm tall with a 270$\times$270 mm footprint and weighs $\sim12$ kg. 

\subsection{Operation} \label{subsec:construction}
\begin{figure}[t]
   \begin{center}
   \begin{tabular}{c}
   \includegraphics[width=0.95\textwidth]{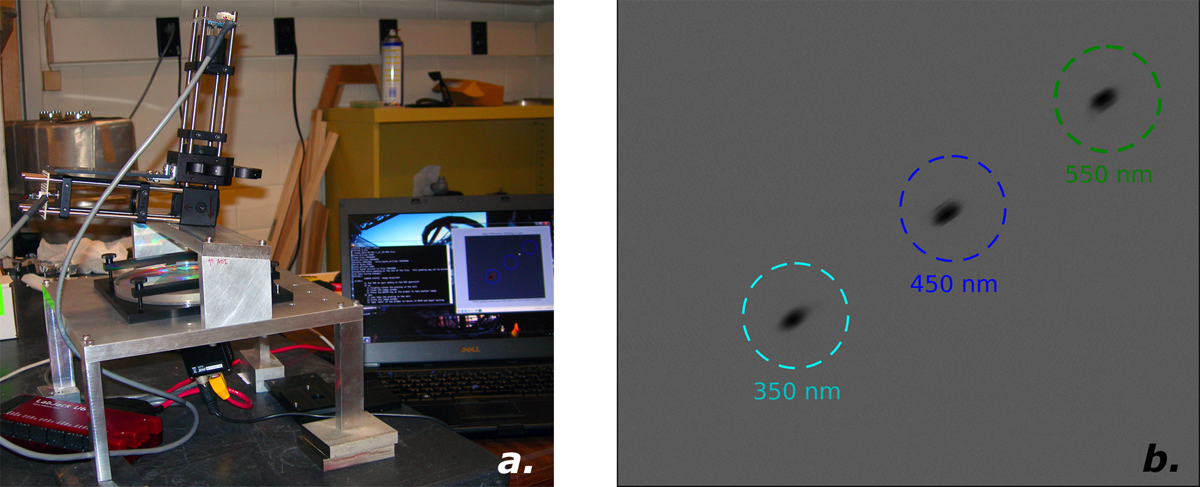}
   \end{tabular}
   \end{center}
   \caption[example] 
   { \label{fig:constructed} 
\textit{a}) A photo of the constructed prototype grating tester. In the background at right is a computer running the command line based control software. \textit{b}) An image constructed from the sum of three dark subtracted CCD images taken with each of the 350, 450 and 550 nm LEDs turned on. Each CCD image was normalized to the peak flux within each indicated circular aperture before summing. The three circles around each image correspond to the photometric apertures used for measurement of the diffraction efficiency at each wavelength. 
}
   \end{figure}

In Fig. \ref{fig:constructed}$a$, we show a photo of the constructed prototype grating tester. For controlling the LED light source and operating and reading the comparison photodiode in photovoltaic mode, we utilize a LabJack U6\footnote{LabJack U6: \textit{http://labjack.com/u6}} data acquisition unit. The U6 requires a connection to a single USB port on the control computer. The native +5 V delivered through the USB interface serves as the power source for the LEDs and the photodiode preamplifier. We have chosen an Allied Vision Technologies Prosilica GC2450\footnote{Allied Vision Technologies Prosilica GC2450: \textit{http://www.alliedvisiontec.com/us/products/cameras/gigabit-ethernet/prosilica-gc/gc2450.html}} CCD camera as our imaging detector. This camera features a 5 megapixel, 2/3''-format sensor with 3.45 $\mu$m square pixels. The camera interfaces with the computer by a Gigabit ethernet port and is powered by an external 12 V DC source. We have verified both the CCD and the comparison photodiode for linearity over the relevant signal levels.

The grating tester is controlled through custom Python software that provides near ``push-button'' simplicity for the tester's operation in a command shell environment. There are two main modes of the software. The first mode is for calibration in which a reference grating is measured whose efficiency $\eta_{\mathrm{ref}}(\lambda)$ is known. Because the diffraction efficiency of a VPH grating can vary as a function of position across the clear aperture, we must take care that the same 12.5mm diameter sub-aperture is used for calibration measurement as was characterized on the bench tester. For this purpose, we have designed a custom cell for sample gratings used for reference that masks off all but the sub-aperture to be measured (see Fig. \ref{fig:gratingcell}$d$). With the reference grating in place, the software begins by taking a set of CCD dark frames to correct for the dark current, which is significant since the GC2450 CCD is uncooled. The user is then prompted to turn the filter wheel (which must be done manually) to the filter corresponding to the first wavelength. The first LED is then turned on, a CCD image is taken, and the photodiode is read out. Both the CCD exposure times and number of readouts of the photodiode signal have been adjusted to give $S/N \approx 100$ for all wavelengths in both device's measurements. The LED is then turned off, and dark measurements are taken with the photodiode to correct for the background. The background subtracted voltage read from the photodiode $V_{\mathrm{ref}}(\lambda)$ is calculated. Using a series of predetermined apertures that are $\sim5\times$ the image size, the total flux $F_{\mathrm{ref}}(\lambda)$ is calculated inside the aperture from the dark subtracted CCD image; the average flux per pixel determined from outside this aperture is used to correct for the background. The values of $F_{\mathrm{ref}}(\lambda)$ and $V_{\mathrm{ref}}(\lambda)$ are saved for later use. This process is repeated for each of the three wavelengths. 

The second mode is for measuring an actual VIRUS grating. The grating is first mounted in its cell and a rough alignment is completed by matching the alignment features on the grating and the cell, as discussed above. A fine alignment is quickly completed by an iterative process in which the user is shown a CCD image with the photometric apertures overlaid (see Fig. \ref{fig:constructed}$b$). The user adjusts the rotation of the grating in the cell by hand, and another CCD image is taken for visual feedback. The user can quit as soon as he/she is satisfied with the alignment and the grating is finally locked down in the cell. The user must then enter the grating's unique vendor-assigned serial number. The test then begins by taking the required CCD dark frames and when complete, the user is prompted to move the grating to the first sub-aperture position and rotate the filter wheel for the first wavelength (note that the choice of which sub-apertures are measured for which wavelengths is user configurable). Once complete, the first LED is turned on and a CCD image and comparison photodiode measurement are taken followed by a photodiode dark measurement. The background corrected flux $F(\lambda)$ and photodiode voltage $V(\lambda)$ are calculated. Finally, the efficiency at that position and wavelength is calculated as:
\begin{equation}\label{eq:testerefficiency}
\eta(\lambda)\: =\: \eta_{\mathrm{ref}}(\lambda)\: \frac{F(\lambda)}{F_{\mathrm{ref}}(\lambda)}\: \frac{V_{\mathrm{ref}}(\lambda)}{V(\lambda)} \; .
\end{equation}
This process is repeated for each wavelength at each sub-aperture defined by the user. Once all wavelengths and sub-apertures have been measured, the software calculates the efficiency for each wavelength averaged over all sub-apertures as well as the total range of measured variation. For useful feedback on the success of the test, the results are immediately plotted and once approved, the data is saved to a user defined file in \texttt{.csv} format. 

Note that while we have not yet implemented an additional automated mode in our software for it, the use of a CCD camera for carrying out these tests gives the opportunity to monitor the scattering properties of the gratings as an additional quality control check. As an example, a grating with particularly bad scattering properties was presented in Fig. 18 of Ref. \citenum{Barden01}. So far, the VIRUS prototype gratings presented in this work appear to have low levels of scattered light. By hand, we have measured $\lesssim2.3$\% of light scattered into apertures placed at 0.5\degree\ on either side of the center of the 350 nm image, perpendicular to the dispersion direction for 327-3, 412-2, and 412-4. Such a measurement will be written into the grating tester control software for automatically monitoring the image quality of the gratings as they are tested for their diffraction efficiency performance.    

\subsection{First Results}\label{subsec:performance}
   \begin{figure}[t]
   \begin{center}
   \begin{tabular}{c}
   \includegraphics[width=0.8\textwidth]{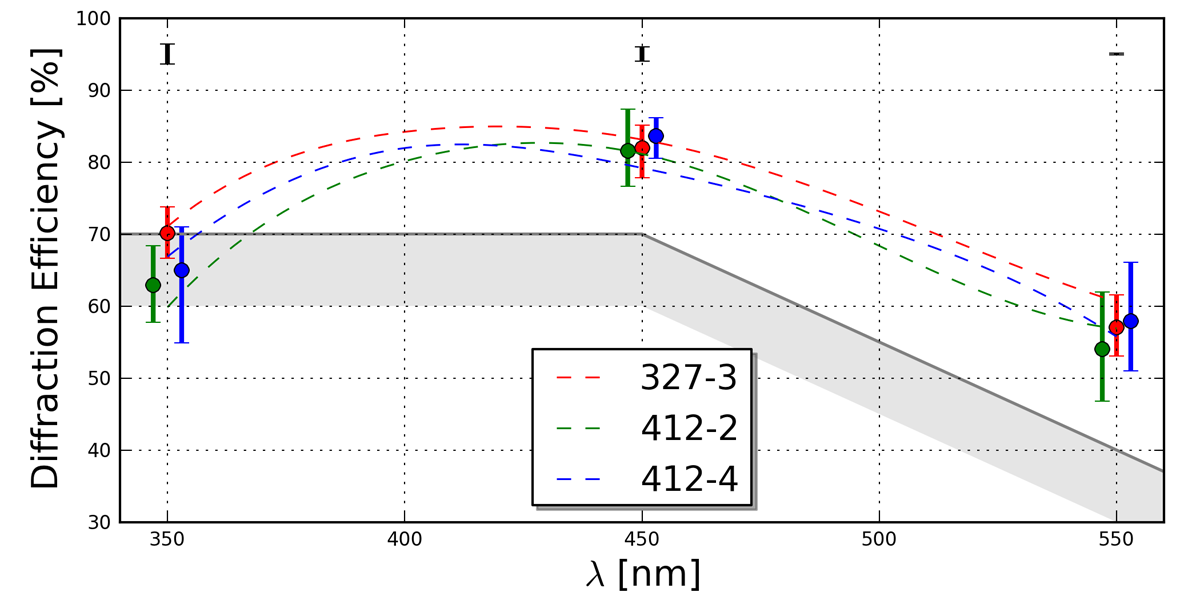}
   \end{tabular}
   \end{center}
   \caption[example] 
   { \label{fig:testerresults} 
First order diffraction efficiency measurements at $\alpha = 9$\degree\ for 327-3, 412-2, and 412-4. The filled circles are data taken with the prototype VIRUS grating tester (averaged over the measurement of nine different 12.5 mm sub-apertures) while for comparison, the dashed curves are cubic spline fits to the bench grating tester data from Fig. \ref{fig:firstorder}$a$ (which were taken for a 12.5 mm sub-aperture whose position on the grating was not recorded). The black error bars above the data indicate the total range of variation in nine measurements of a \textit{single} sub-aperture, which reflects the photometric uncertainty. The colored error bars correspond to the total range of variation measured from nine \textit{different} sub-apertures, which reflects the spatial variability of the diffraction efficiency. Note that the data points have been offset by 3 nm from each other at each wavelength for clarity.  
}
   \end{figure} 
To prove the concept of the VIRUS grating tester, we have performed full tests for all three wavelengths and all nine sub-apertures of 327-3, 412-2, and 412-4. A full efficiency test from beginning to end (including alignment and setup) takes $\sim10$ minutes to complete. The results are shown in Fig. \ref{fig:testerresults} for all three gratings. We confirm, as shown in $\S$\ref{subsec:1storder}, that the gratings perform comparably for $m=1$ and $\alpha=9$\degree, with 327-3 performing slightly better at 350 nm than the other two. The agreement with the bench test facility within the measured spatial variation of the diffraction efficiency successfully proves the concept of the VIRUS grating tester. A given measurement is extremely repeatable thanks to the fixed mechanical design and the high $S/N$ that we can achieve. 

These measurements also highlight the importance of averaging the efficiency over multiple sub-apertures spread over the clear aperture for physically large gratings. Spatial variation of the diffraction efficiency for a VPH grating can be caused by several imperfections in the fabrication process, such as non-uniform processing of the DCG and aberrations in the optics used to create the interference pattern to which the grating layer is exposed. With smaller and fewer sub-apertures (such as a single efficiency measurement made with a laser, a test that is often used by grating vendors to verify that they have met an order's specification), one is more prone to over or underestimating a grating's characteristic diffraction efficiency by measuring a particular area with above or below average efficiency. Ideally, one would measure the efficiency with a collimated beam matched to the grating's clear aperture. For gratings such as ours with a fairly large diameter, such a test apparatus would be overly large and expensive to build. As our results show, it is not uncommon to have a VPH grating with spatial efficiency variations of up to $\pm8$\% when measured with even a 12.5 mm diameter beam. These spatial variations are real, as they are a factor of at least a few greater than the photometric uncertainties that are determined by repeatedly measuring the same sub-aperture for each wavelength. This level of spatial variation should be considered when looking at the results presented in $\S$\ref{subsec:1storder} and $\S$\ref{subsec:ordertests}, which were for a single 12.5 mm diameter sub-aperture. In general, the spatially averaged diffraction efficiency of the three gratings shows that they are virtually indistinguishable at 450 and 550 nm, which supports our conclusion from $\S$\ref{subsec:1storder} that the fabrication of these VPH gratings can be done consistently enough for our requirements. We note, however, that the spatially averaged diffraction efficiencies also suggest that the below batch mean efficiencies of 412-2 and 412-4 cannot be blamed alone on having by chance measured a below average sub-aperture in our original single sub-aperture tests. 

\section{CONCLUSIONS} \label{sec:conclusions}
We have presented the design of the VPH diffraction gratings that will be utilized in the new VIRUS array of spectrographs for the HET. We have fabricated and extensively tested four prototype, production-grade VPH gratings. Our tests have shown that we are very near to meeting the external diffraction efficiency that is required for the HETDEX survey to meet its goals. Additionally, our tests have identified various discrepancies between the data and RCWA models for our particular grating prescription. We are currently unable to explain these discrepancies. By measuring various diffraction orders and attempting to take into account the transmission of the grating layer, substrates, and optical epoxy, we have identified some issues with the DCG layer in our gratings that contribute to increased scattering and a gradual drop in performance towards 350 nm. We are currently working to improve the fabrication process, which should lead to improvement in the overall external diffraction efficiency. Despite some of the issues that our tests have uncovered, we have proposed very slight design changes that will allow gratings with the current prescription and fabrication process to meet the VIRUS batch mean diffraction efficiency requirement. By the end of summer 2012, we expect the suite of $\sim200$ VIRUS gratings to be in production.

With such a large number of gratings, we are faced with the challenge of verifying each of their performance for acceptance. We have thus presented the design and verified the concept of the VIRUS grating tester, a simple apparatus that will allow a full characterization of the $m=1$ external diffraction efficiency averaged over the grating clear aperture for the designed angle of incidence in $\sim10$ minutes per grating. We are currently purchasing and fabricating the required parts for a production version of the tester. The production tester and a dedicated computer will be shipped to the vendor when the contract is awarded. Vendor personnel will be trained to operate the tester, and the results will be sent to us as each of the grating batches are complete. A future paper will describe the performance of the final production suite of gratings.

\acknowledgments     
HETDEX is run by the University of Texas at Austin McDonald Observatory and Department of Astronomy with participation from the Ludwig-Maximilians-Universit\"{a}t M\"{u}nchen, Max-Planck-Institut f\"{u}r Extraterrestriche-Physik (MPE), Leibniz-Institut f\"{u}r Astrophysik Potsdam (AIP), Texas A\&M University, Pennsylvania State University, Institut f\"{u}r Astrophysik G\"{o}ttingen, University of Oxford, and Max-Planck-Institut f\"{u}r Astrophysik (MPA).  In addition to Institutional support, HETDEX is funded by the National Science Foundation (grant AST-0926815), the State of Texas, the US Air Force (AFRL FA9451-04-2-0355), and generous support from private individuals and foundations. 

Thanks to Chad Mosby for helping to fabricate the prototype VPH gratings. We recognize the following individuals for their contributions to the design and fabrication of components for the VIRUS Grating Tester: N. Canac, D. Edmonston, L. Fuller, M. Rafal, T. Rafferty, C. Ramiller,  R. Ruiz, T. Taylor, and B. Vattiat.

\bibliography{report}         
\bibliographystyle{spiebib}   

\end{document}